\documentclass[superscriptaddress, nofootinbib, prd]{revtex4}[12pt]
\usepackage{graphicx,bm}
\graphicspath{{figs/}}
\usepackage{amsmath,amssymb,amsfonts,rotating}
\usepackage{hyperref}
\usepackage[usenames,dvipsnames]{color}
\usepackage{mdwlist} 
\usepackage{slashed}
\usepackage{verbatim}
\usepackage{mathrsfs}
\usepackage[english]{babel}

\begin{document}

\def\cL{{\cal L}}
\def\be{\begin{eqnarray}}
\def\ee{\end{eqnarray}}
\def\bea{\begin{eqnarray}}
\def\eea{\end{eqnarray}}
\def\beq{\begin{eqnarray}}
\def\eeq{\end{eqnarray}}
\def\tr{{\rm tr}\, }
\def\nn{\nonumber\\}
\def\e{{\rm e}}


\title{Covariant Ho\v{r}ava-like and mimetic Horndeski gravity:\\cosmological solutions and perturbations}

\author{Guido Cognola\footnote{Email address: guido.cognola@unitn.it}}
\affiliation{Department of Physics, University of Trento,\\Via Sommarive 14, I-38123 Povo (TN), Italy}
\affiliation{Istituto Nazionale di Fisica Nucleare (INFN), Gruppo Collegato di Trento,\\Via Belenzani 12, I-38050 Povo (TN), Italy}
\affiliation{Trento Institute for Fundamental Physics and Applications (TIFPA),\\Via Sommarive 14, I-38123 Povo (TN), Italy}

\author{Ratbay Myrzakulov\footnote{Email address: rmyrzakulov@gmail.com}}
\affiliation{Department of General \& Theoretical Physics and Eurasian Center for
Theoretical Physics,\\Eurasian National University, 010008 Astana, Satpayev Str. 2, Kazakhstan}

\author{Lorenzo Sebastiani\footnote{Email address: lorenzo.sebastiani@unitn.it}}
\affiliation{Department of General \& Theoretical Physics and Eurasian Center for
Theoretical Physics,\\Eurasian National University, 010008 Astana, Satpayev Str. 2, Kazakhstan}

\author{Sunny Vagnozzi\footnote{Email address: sunny.vagnozzi@fysik.su.se}}
\affiliation{The Oskar Klein Centre for Cosmoparticle Physics, Department of Physics, Stockholm University, Albanova, SE-106 91 Stockholm, Sweden}
\affiliation{NORDITA (Nordic Institute for Theoretical Physics), KTH Royal Institute of Technology and Stockholm University, Roslagstullbacken 23, SE-106 91 Stockholm, Sweden}

\author{Sergio Zerbini\footnote{Email address: sergio.zerbini@unitn.it}}
\affiliation{Department of Physics, University of Trento,\\Via Sommarive 14, I-38123 Povo (TN), Italy}
\affiliation{Istituto Nazionale di Fisica Nucleare (INFN), Gruppo Collegato di Trento,\\Via Belenzani 12, I-38050 Povo (TN), Italy}
\affiliation{Trento Institute for Fundamental Physics and Applications (TIFPA),\\Via Sommarive 14, I-38123 Povo (TN), Italy}


\begin{abstract}
We consider a variant of the Nojiri-Odintsov covariant Ho\v{r}ava-like gravitational model, where diffeomorphism invariance is broken dynamically via a non-standard coupling to a perfect fluid. The theory allows to address some of the potential instability problems present in Ho\v{r}ava-Lifshitz gravity due to explicit diffeomorphism invariance breaking. The fluid is instead constructed from a scalar field constrained by a Lagrange multiplier. In fact, the Lagrange multiplier construction allows for an extension of the Ho\v{r}ava-like model to include the scalar field of mimetic gravity, an extension which we thoroughly explore. By adding a potential for the scalar field, we show how one can reproduce a number of interesting cosmological scenarios. We then turn to the study of perturbations around a flat FLRW background, showing that the fluid in question behaves as an irrotational fluid, with zero sound speed. To address this problem, we consider a modified version of the theory, adding higher derivative terms in a way which brings us beyond the Horndeski framework. We compute the sound speed in this modified higher order mimetic Ho\v{r}ava-like model and show that it is non-zero, which means that perturbations therein can be sensibly defined. Caveats to our analysis, as well as comparisons to projectable Ho\v{r}ava-Lifshitz gravity, are also discussed. In conclusion, we present a theory of gravity which preserves diffeomorphism invariance at the level of the action but breaks it dynamically in the UV, reduces to General Relativity in the IR, allows the realization of a number of interesting cosmological scenarios, is well defined when considering perturbations around a flat FLRW background, and features cosmological dark matter emerging as an integration constant.
\end{abstract}

\pacs{95.36.+x, 98.80.Cq}

\maketitle

\section{Introduction}
\label{introduction}

Of the four fundamental interactions, gravity retains a special status in that the construction of the corresponding quantum theory has thus far proven elusive, due to the non-renormalizability of General Relativity (GR). It is clear that GR should be viewed as an effective field theory (EFT), bound to break down at some high energy scale (presumably the Planck scale), and which contains only the leading term in a curvature expansion. One possibility to render gravity renormalizable is to modify the UV behaviour of the graviton propagator, for instance by adding higher order curvature terms. While it is known that actions constructed from invariants quadratic in curvature are renormalizable \cite{stelle}, the addition of higher order curvature terms in fact correspond to the addition of higher order time derivatives, which lead to the appearance of ghost degrees of freedom and hence a loss of unitarity (see e.g. \cite{bos} for a general review). Clearly, then, a better solution would be that of improving the UV behaviour of the graviton propagator by adding higher order spatial but not time derivatives, which in turn requires treating space and time on a different footing, leading to Lorentz violation. The use of Lorentz violation in the UV as a field theory regulator is a possibility which had already been considered in the past, although the amplitude of Lorentz violation should be appropriately suppressed in the IR. In particular, the formulation of non-relativistic theories of gravity has been driven by endeavours to describe non-relativistic field theories via AdS/CFT. \\

Ho\v{r}ava-Lifshitz gravity (HLG hereafter) \cite{horava}, proposed by Petr Ho\v{r}ava in 2009, is an attempt to embed these heuristic ideas within a rigorous framework. The idea behind the proposal is deceivingly simple: to make gravity power-counting renormalizable by abandoning Lorentz symmetry in favour of a Lifshitz-type anisotropic scaling in the UV. The theory is compatible with anisotropic scaling of the space and time coordinates ($\mathbf{x}$ and $t$ respectively) with dynamical critical exponent $z$, that is:
\begin{eqnarray}
t \rightarrow b^zt \, , \quad \mathbf{x} \rightarrow b\mathbf{x} \, .
\label{horava}
\end{eqnarray}
The above should be understood in the following way: the theory possesses a solution which describes an UV fixed point with scaling properties described by Eq.(\ref{horava}). If anisotropic scaling with $z \geq 3$ is realized in the UV, the theory is power-counting renormalizable in 3+1 dimensions, which leads to its conjectured renormalizability. Moreover, the theory naturally flows to $z = 1$ in the IR. It is clear that for $z \neq 1$, Lorentz invariance is lost.\footnote{In fact, in the context of a more fundamental theory where spacetime itself is emergent, it is difficult to conceive how Lorentz invariance could be preserved at a fundamental level.} The theory is instead invariant under space-independent time reparametrizations and time-dependent spatial diffeomorphisms, that is, transformations of the form:
\begin{eqnarray}
t \rightarrow t'(t) \, , \quad \mathbf{x} \rightarrow \tilde{\mathbf{x}}(t,\mathbf{x}) \, .
\label{fpd}
\end{eqnarray}
The map in Eq.(\ref{fpd}) is known as a foliation preserving diffeomorphism. In HLG, unlike GR, the foliation of spacetime by constant hypersurfaces is therefore more than just a mere choice of coordinates. For a recent comprehensive review on HLG, see for instance \cite{sotiriou1}, whereas \cite{unifiedno} deals with $F(R)$ HLG. We instead refer the reader to \cite{calcagni,kofinas,brandenberger,mukohyama1,sotiriou2,mukohyama2} for discussions on cosmology within HLG.\footnote{See also \cite{saridakis1} for a recent work on the emergent Universe scenario within HL F(R) cosmology.} \\

Despite its many successes, HLG has been at the receiving end of criticisms in a number of works (see e.g. \cite{charmousis,li,blas,blas1,blas2,blas3} for some of the early criticisms). It has been argued that the theory possesses additional unphysical modes, associated to the explicit breaking of diffeomorphism invariance. Such modes, it is argued, do not decouple and actually become strongly coupled in the IR, preventing the recovery of the perturbative GR limit at low energies. A particular solution to this problem was presented in 2009, when Nojiri and Odintsov constructed a Ho\v{r}ava-like model retaining full diffeomorphism invariance at the level of the action \cite{NO1}. In the following, we will refer to such model as covariant renormalizable gravity (CRG henceforth). The model features a non-standard coupling of curvature to the energy-momentum tensor of an exotic perfect fluid. When considering perturbations around the background, diffeomorphism invariance is broken dynamically by this non-standard coupling. Due to diffeomorphism invariance being preserved at the level of the action, only physical transverse modes propagate. As in the case of HLG, the graviton propagator behaves as $1/\mathbf{k}^{2z}$ in the UV, where $\mathbf{k}$ denotes spatial momenta and $z \geq 3$ is required in order for the theory to be power-counting (super-)renormalizable. Moreover, a consistent theory was constructed for any integer value of $z$. The model was reformulated and generalized in \cite{NO2}, where an effective fluid with arbitrary equation of state (EoS) parameter $w$ was constructed by means of a Lagrange multiplier constrained scalar field, following ideas first presented in \cite{lim,gao,capozziello} to unify dark matter and dark energy (see also \cite{paliathanasis}). Following the proposal by Nojiri and Odintsov, other theories with similar properties to the original CRG model were proposed and studied (e.g. \cite{cognola}), so that it is actually more appropriate to view such theories as a more general class of Ho\v{r}ava-like models which dynamically obtain a preferred foliation. \\

The price to pay in CRG-like models is the presence of the aforementioned exotic fluid, which does not correspond to the usual perfect fluids found in cosmology. The origin of this fluid is unknown, although it could be string-inspired. However, it was recently realized in \cite{comment} that it is possible to connect CRG-like models to mimetic gravity, as the two theories both feature a fluid realized through a Lagrange multiplier constrained term. Recall that in mimetic gravity (to be discussed more thoroughly later), one isolates the conformal degree of freedom of gravity in a covariant way, by parametrizing the physical metric in terms of an auxiliary metric and a scalar field (the mimetic field). The Lagrange multiplier term in the action constrains the 4-gradient of the mimetic field, which in turn induces an effective fluid, whose behaviour mimics that of collisionless cold dark matter. This suggests that it is possible to extend CRG-like models by including the mimetic field. Mimetic gravity is also a special case of the most general second-order scalar-tensor theory of gravity known as Horndeski gravity \cite{horndeski}, and is related to GR by singular disformal transformations. In fact, recently a general scalar-tensor mimetic Horndeski theory has been considered, and it has been argued that the two approaches to mimetic gravity (i.e. disformal transformations and Lagrange multiplier constraints) are indeed equivalent. Motivated by the recent spur of interest towards Ho\v{r}ava-Lifshitz gravity, mimetic gravity, and Horndeski gravity, our aim in this work is to rigorously explore the connections between these different theories, and in particular to CRG-like models. In particular, it is our objective to inspect the consequences of extending a variant of the original covariant renormalizable gravity model of Nojiri and Odintsov to include the mimetic field. \\

The paper is structured as follows. Having reviewed the basic ideas of Ho\v{r}ava-Lifshitz gravity in Section \ref{introduction}, we will proceed to briefly review CRG-like models and mimetic gravity in Section \ref{crg}, outlining the connections between the two. Section \ref{gcrg} will be devoted to generalizing a specific CRG-like model by including the mimetic field and correspondingly a potential for the latter. In Section \ref{solutions} we will explore some cosmological solutions of the mimetic CRG-like model. In the same section we will also consider perturbations around a flat FLRW background, and note that the speed of sound therein is identically zero, thus preventing us from defining perturbations in the mimetic field in the usual way. To overcome this drawback, in Section \ref{modified} we consider a modified generalized version of our original mimetic CRG-like model, by adding higher derivative terms. We show that in the modified version, the sound speed is nonvanishing. We conclude in Section \ref{conclusion} by reviewing our work and providing final remarks. Throughout the paper, we will set $8\pi M_{Pl}^2=1$, where $M_{Pl}$ is the Planck mass. \\

\section{Covariant renormalizable gravity-like models and mimetic gravity}
\label{crg}

\subsection{Covariant renormalizable gravity-like models}

Let us first review covariant renormalizable gravity-like models, and then focus on one particular such model. The original CRG action proposed by Nojiri and Odintsov in \cite{NO1} takes the form:
\begin{eqnarray}
S = \int d^4x \ \sqrt{-g} \left [ \frac{R}{2} - \alpha \left ( \tau ^{\mu \nu}R _{\mu \nu} + \beta \tau R \right ) ^2 \right ] \, , \quad \beta = \frac{w-1}{2(1 - 3w)} \, .
\label{action}
\end{eqnarray}
In the above, $\tau _{\mu \nu}$ is the energy-momentum tensor of the perfect fluid, whose EoS parameter is $w$. When perturbing around a flat background, the term in the action [Eq.(\ref{action})] proportional to $\alpha$ will only contain spatial and not time derivatives, and hence breaks full diffeomorphism invariance dynamically. It can be shown that the graviton propagator goes as $1/\mathbf{k}^4$ in the UV: that is, we have recovered a dynamical $z = 2$ Ho\v{r}ava-like theory. The argument holds if $w \neq -1,1/3$ for in the former case the coupling to the fluid vanishes and hence the graviton propagator behaviour is not modified, whereas in the latter the coupling itself diverges. As was shown in \cite{NO1}, due to the absence of explicit symmetry breaking terms, only transverse physical modes propagate in the theory, unlike the case of HLG where unphysical longitudinal modes which are strongly coupled in the IR might appear. The above arguments hold in a curved background as well \cite{NO1}. \\

The action in Eq.(\ref{action}) gave us the $z = 2$ Ho\v{r}ava Lifshitz-like theory, but can be easily generalized to accommodate any $z \geq 3$. The general action given by:
\begin{eqnarray}
S = \int d^4x \ \sqrt{-g} \left \{ \frac{R}{2} - \alpha \left [ \left ( \tau^{\mu \nu}\nabla _{\mu}\nabla _{\nu} + \gamma \tau\nabla ^{\rho}\nabla _{\rho} \right ) ^n \left ( \tau^{\mu \nu} R _{\mu \nu} + \beta \tau R \right ) \right ] ^2 \right \} \, , \quad \beta = \frac{w-1}{2(1 - 3w)} \, , \quad \gamma = \frac{1}{3w - 1} \, .
\label{crg1}
\end{eqnarray}
which corresponds to the $z = 2n + 2$ Ho\v{r}ava-like theory. For odd $z$, one may make use of two copies of
 $\left( \tau ^{\mu \nu}R _{\mu \nu} + \beta \tau R \right ) $ on either side of the differential operator. Clearly, when $n = 0$, we recover the basic action in Eq.(\ref{action}). The terms inducing dynamical Lorentz symmetry breaking contain higher derivatives and hence are relevant only in the UV region. In the IR limit GR is recovered. \\

The model was reformulated in \cite{NO2} following ideas presented in \cite{lim,gao,capozziello}. The idea is to construct the fluid with arbitrary EoS parameter $w$ from a scalar field satifsying a constraint enforced through a Lagrange multiplier term. Because of the constraint, the scalar field can be non-dynamical and even in the high energy region one can obtain a non-relativistic fluid, as required in CRG. The Nojiri-Odintsov action for $z = 2n+2$ CRG formulated with a Lagrange multiplier (which we will refer to as LCRG) reads:
\begin{eqnarray}
S = \int d^4x \ \sqrt{-g} \left \{ \frac{R}{2} - \alpha \left [ \left ( \partial ^{\mu}\phi\partial ^{\nu}\phi \nabla _{\mu}\nabla _{\nu} + 2U_0\nabla ^{\rho}\nabla _{\rho} \right ) ^n \left ( \partial ^{\mu}\phi\partial ^{\nu}\phi R_{\mu \nu} + U_0R \right ) \right ] ^2 - \lambda \left ( \frac{1}{2}\partial _{\mu}\phi \partial ^{\mu}\phi + U_0 \right ) \right \} \, .
\label{lcrg}
\end{eqnarray}
As expected, the Nojiri-Odintsov action, which is diffeomorphism invariant, is now expressed entirely in terms of local fields. Again, given a certain $n$, the graviton propagator behaves as $1/\mathbf{k}^{2n+2}$ in the UV, thus leading to (super-)renormalizability in 3+1 dimensions if $z = (\geq) 3$. In \cite{kluson1,tureanu1,kluson2,saezgomez,kluson3}, the $F(R)$ gravity version of Ho\v{r}ava gravity has been formulated, whereas the $F(R)$ version of CRG and LCRG have been constructed in \cite{carloni} and \cite{NO2} respectively, and are straightforward generalizations of Eqs.(\ref{crg1},\ref{lcrg}). See also \cite{NO3} for further generalizations of CRG. In the IR, all these theories reduce to GR and have been shown to recover Newton's law. \\

As we mentioned earlier, other CRG-like models (which break diffeomorphism invariance dynamically by non-standard coupling to a fluid) have been proposed in the literature. In particular, Cognola et al. \cite{cognola} considered a model defined by the following action:
\begin{eqnarray}
I = \frac{1}{2}\int d ^4x\sqrt{-g} \left \{ R - 2\Lambda - \alpha \left [ \left ( R ^{\mu \nu} - \frac{R}{2}g ^{\mu \nu} \right ) \nabla _{\mu}\phi\nabla _{\nu}\phi \right ] ^n - \frac{\lambda}{2} \left ( g ^{\mu \nu}\nabla _{\mu}\phi\nabla _{\nu}\phi + 1 \right ) \right \} \ ,
\label{cognolaa}
\end{eqnarray}
and studied black hole and cosmological solutions therein. The non-minimal coupling considered in Eq.(\ref{cognolaa}) is actually not new. Couplings similar to this Horndeski-like one (for $n = 1$) have been considered in the literature (refer for instance to \cite{rinaldi}), particularly in connection to UV-protected natural inflation \cite{germani,watanabe}. Whereas the original CRG model has been the subject of much further study, the CRG-like model of Cognola et al. remains essentially unexplored. For this reason, in our work we choose to focus on such model, exploring in particular connections and extensions to mimetic gravity, cosmological solutions, and the behaviour of scalar perturbations in such theory.

\subsection{Mimetic gravity}

Fluids constructed from scalar fields constrained by a Lagrange multiplier abound in the recent literature. For instance, this construction is at the heart of the mimetic gravity framework. In mimetic gravity one isolates the conformal degree of freedom of gravity in a covariant way, by parametrizing the physical metric $g$ in terms of an auxiliary metric $\tilde{g}$ and the mimetic field $\phi$ as follows \cite{mukhanov1}:
\begin{eqnarray}
g _{\mu \nu} = -\tilde{g} _{\mu \nu}\tilde{g} ^{\alpha \beta}\partial _{\alpha}\phi \partial _{\beta}\phi \, .
\end{eqnarray}
The corresponding equations for the gravitational field are equivalent to the Einstein field equations, with the addition of a source term, which describes a pressureless fluid with 4-velocity $\partial _{\mu}\phi$. This additional degree of freedom can mimic collisionless cold dark matter. Of course, the following equality has to hold for consistency:
\begin{eqnarray}
g ^{\mu \nu}\partial _{\mu}\phi \partial _{\nu}\phi = -1 \, .
\label{constraint}
\end{eqnarray}
In fact, this suggests that the constraint given by Eq.(\ref{constraint}) can be implemented in the action by means of a Lagrange multiplier \cite{mukhanov2}. In other words, the action for mimetic gravity can be written as:
\begin{eqnarray}
S = \int d^4x \ \sqrt{-g} \left [ \frac{R}{2} - \lambda \left ( g ^{\mu \nu}\partial _{\mu}\phi \partial _{\nu}\phi + 1 \right ) -V(\phi) \right ] \, ,
\label{mimetic}
\end{eqnarray}
where $V(\phi)$ represents a potential for the mimetic field.\\

It became soon understood that mimetic gravity is related to GR via a singular disformal transformation. Recall that, by virtue of diffeomorphism invariance of GR, any metric can be parametrized in terms of a fiducial metric and a scalar field \cite{bekenstein}. If the transformation is invertible, the number of degrees of freedom is unchanged by the transfomation and one recovers GR. If, as is the case in mimetic gravity, the transformation is singular, then the number of degrees of freedom can change and one has, in general, equations of motion which differ from those of GR \cite{deruelle,domenech1,domenech2}. More recently still, \cite{arroja} has shown how the two approaches to mimetic gravity, that is, singular disformal transformations and Lagrange multiplier, are in fact equivalent.\footnote{See also \cite{ghalee1,carvalho,ghalee2} for recent works on the role of disformal transformations in cosmology} Mimetic gravity has received a tremendous amount of interest over the past years, with several extensions and solutions being formulated and derived, and various studies conducted on the stability against negative energy states. It is beyond the scope of this section to review in detail findings concerning mimetic gravity, for which instead we refer the reader to the rapidly developing literature on the subject \cite{golovnev,barvinsky,myrzakulov1,oksanen1,malaeb,haghani1,myrzakulov2,odintsov1,
saadi,capela,vikman,saridakis,haghani2,odintsov2,yuan,myrzakulov3,myrzakulov4,odintsov3,myrzakulov5,oksanen2,
silva,myrzakulov6,myrzakulov7,deffayet,khalifeh,ramazanov,kunz,haghani3,myrzakulov8,guendelman1,kan,odintsov4,
guendelman2,rabochaya,oikonomou1,nurgissa,hojman,myrzakulov9,odintsov5,guendelman3,oikonomou2,myrzakulov10,
guendelman4,koshelev,odintsov6,ali,hammer,odintsov7,bartolo,myrzakulov11,inhomogeneousjcap,pilo,odintsov8,babichev,
odintsov9,biscalar,langlois,zumalacarregui,weiner,suyama,saitou,sepangi,steinhardt,oikonomou3,kopp,quanta,
vacaru,odintsov10}. \\

Early on it was realized that the original mimetic gravity proposal suffers a serious problem, namely that the perturbations in the mimetic field behave as dust with a speed of sound which is identically zero irrespective of wavelength \cite{mukhanov2}. That is, there is no dependence on the Laplacian of the perturbation in the perturbation equation itself. This is unacceptable, as it implies that it is not possible to define quantum perturbations in mimetic matter as one would usually do, else these would fail to seed the observed large-scale structure in the Universe. This is of course not unexpected, given identical results obtained in \cite{lim}. Parallel work has demonstrated how this is the case in the most generic mimetic Horndeski theory of gravity \cite{bartolo}. This can easily be understood: the Lagrange multiplier constraint kills the wave-like parts of the would-be scalar degree of freedom, and hence there can be no propagating scalar degree of freedom. A possible solution is the addition of higher derivative (HD) terms, which modify the sound speed and even have the potential to address some of the outstanding problems of the standard collisionless cold dark matter picture on small scales \cite{capela,vikman,ramazanov}. \\

Because CRG-like models and mimetic gravity both feature fluids constructed from Lagrange multipliers [see the actions given by Eqs.(\ref{cognolaa},\ref{mimetic})], it is possible to extend the CRG-like model to include the mimetic field, and correspondingly a potential for it. It is therefore our goal to explore the consequences of this extension, and comment on connections of such theory to other healthy extensions of Ho\v{r}ava gravity. Our purpose in this work is twofold. First, having identified the connection between CRG-like models and mimetic gravity, we generalize the former by adding a potential for the mimetic field, and examine whether this theory features interesting cosmological solutions. \\

Second, we study perturbations of such model around a flat FLRW background, and notice that in doing so the resulting sound speed is identically zero, which renders the quantization of the theory problematic. We solve this problem by modifying the theory by the addition of higher derivative terms, and show that it is necessary to go beyond the Horndeski framework in order to have a non-zero sound speed. This is not unexpected, given that \cite{bartolo} has shown in all generality that the sound speed in mimetic Horndeski models, within which our modified covariant Ho\v{r}ava-like model falls, is zero. The end product of our study is a covariant Ho\v{r}ava-like theory of gravity which preserves diffeomorphism invariance at the level of the action (thus circumventing the possible strong-couling IR instability issues of HLG) but breaks it dynamically in the UV, reduces to GR in the IR, allows the realization of a number of interesting cosmological scenarios, has cosmological dark matter emerging as an integration constant, and is well defined when considering perturbations which will seed the observed large-scale structure. \\

\section{Generalized Nojiri-Odintsov covariant gravity}
\label{gcrg}

In the light of the connection identified between CRG-like models and mimetic gravity, we extend the action of the CRG-like model constructed by Cognola et al., given by Eq.(\ref{cognolaa}), by adding a potential for the mimetic field, such field inducing the exotic fluid coupling to curvature. The action of the theory is therefore given by:
\begin{eqnarray}
I = \frac{1}{2}\int d ^4x\sqrt{-g} \left \{ R - 2\Lambda - \alpha \left [ \left ( R ^{\mu \nu} - \frac{R}{2}g ^{\mu \nu} \right ) \nabla _{\mu}\phi\nabla _{\nu}\phi \right ] ^n - \frac{\lambda}{2} \left ( g ^{\mu \nu}\nabla _{\mu}\phi\nabla _{\nu}\phi + 1 \right )  - V (\phi) \right \} \ ,
\label{actionpotential}
\end{eqnarray}
where $g$ is the determinant of the metric tensor $g_{\mu \nu}$, $R_{\mu \nu}$ and $R$ are the Ricci tensor and scalar respectively, $n$ is a positive integer, $\phi$ is the mimetic field with potential $V(\phi)$, $\Lambda$ is the cosmological constant which can be incorporated in $V(\phi)$, and $\nabla_{\mu}$ denotes the covariant derivative (see also \cite{NO3}).\footnote{Note that $\nabla_{\mu} \phi \equiv \partial_{\mu} \phi$, being $\phi$ a scalar field.} A theory with the action given by Eq.(\ref{actionpotential}) may also be seen as an example of $F(R,T,R_{\mu \nu}T^{\mu \nu})$ gravity, see \cite{saezgomez1}. \\

Two important comments are in order. First, when the potential in Eq.(\ref{actionpotential}) is a constant, the model can actually be identified with previously studied theories. More precisely, we can relate it to the khronon formalism of Ho\v{r}ava-Lifshitz gravity \cite{blas3}. The only difference with respect to such theories lies in the fact that, through the constraint enforced by the Lagrange multiplier, the unit vector which appears in the khronon framework can be identified with the gradient of a scalar field. After gauge fixing, this can be shown to imply that $g_{00}$ is constant on spatial slices, i.e. the lapse function in Ho\v{r}ava-Lifshitz gravity is a function of time only. This corresponds to the condition for projectable Ho\v{r}ava-Lifshitz gravity.\footnote{We thank the referee for bringing this point to our attention.} We further notice that such identification should not come as a surprise, given that the model we are dealing with is an extension of mimetic gravity. In fact, several works \cite{haghani1,capela,vikman,myrzakulov5,hammer,steinhardt} had previously discussed the fact that mimetic gravity is related to the scalar Einstein-aether theory, where the aether vector (which itself is related to the khronon field, as both define a preferred time) is restricted to being the gradient of a scalar field. In \cite{jacobson}, it was determined that, when the potential for the scalar field is a constant, the model is equivalent to the IR limit of projectable Ho\v{r}ava-Lifshitz gravity, which thus confirms this important identification.\footnote{Furthermore, a formal proof of the equivalence between mimetic gravity and the IR limit of projectable Ho\v{r}ava-Lifshitz gravity has been provided in \cite{pilo}.} We will further comment on the correspondence between our CRG-like mimetic Horndeski model and the projectable version of Ho\v{r}ava-Lifshitz gravity in Section V, when we will modify the model by adding higher derivative terms. \\

A second comment refers to the fact that, when the potential is a constant, the mimetic scalar field enjoys a shift symmetry. When a non-constant potential is included, the original shift symmetry is broken. Then, the renormalization group flow could generate terms which are relevant with respect to the terms in the action given by Eq.(\ref{actionpotential}), and which could thwart the renormalizability of the model. This issue is very important, and generally affects modified mimetic theories. While the problem absolutely deserves further inspection into, it also goes beyond the scope of the present work, which is the study of solutions and perturbations within this theory. Therefore, we postpone the study of the (non)-renormalizability of the CRG-like mimetic model, as well as possible solutions to this problem, to a future work. \\

We now derive the equations of motion of the theory defined by the action Eq.(\ref{actionpotential}). These are obtained by varying the action with respect to Lagrange multiplier, mimetic field, and metric. Variation with respect to the Lagrange multiplier $\lambda$ yields the constraint equation on the scalar field:
\begin{eqnarray}
g^{\mu \nu} \nabla_{\mu} \phi \nabla_{\nu} \phi = -1 \, .
\label{multiplierequation}
\end{eqnarray}
Variation with respect to the scalar field leads to the following equation of motion:
\begin{eqnarray}
V'(\phi) & = & \nabla _{\mu} \left [ \left ( 2n\alpha F ^{n-1} G ^{\mu \nu} + \lambda g ^{\mu \nu} \right ) \partial _{\nu}\phi  \right ] \nonumber\\
& = & \frac{1}{\sqrt{-g}}\partial _{\mu} \left \{ \sqrt{-g} \left [ \left ( 2n\alpha F ^{n-1}G ^{\mu \nu} + \lambda g ^{\mu \nu} \right ) \partial _{\nu}\phi \right ] \right \} \, ,
\label{equationscalar}
\end{eqnarray}
where $V'(\phi)$ indicates the derivative of the potential respect to the field, $G ^{\mu \nu} \equiv R ^{\mu \nu} - g ^{\mu \nu}R/2$, and we have defined:
\begin{eqnarray}
F \equiv T_{\mu \nu}R^{\mu \nu}-\frac{RT}{2} \, , \qquad T_{\mu \nu} \equiv \nabla_{\mu} \phi \nabla_{\nu} \phi \, , \qquad T \equiv g^{\mu \nu}T_{\mu \nu}=-1 \, .
\end{eqnarray}
Finally, by varying with respect to the metric we obtain the equations for the gravitational field (see also \cite{cognola}):
\begin{eqnarray}
G_{\mu \nu}+\Lambda g_{\mu \nu} + \frac{\alpha}{2}F^ng_{\mu \nu} & = & n\alpha F^{n-1} \left [ R^{\rho}_{\mu}T_{\rho \nu} + R^{\rho}_{\nu}T_{\rho \mu} - \frac{1}{2} \left ( TR_{\mu \nu} + RT_{\mu \nu} \right ) \right ] + \frac{\lambda}{2}T_{\mu \nu} \nonumber \\
& + & n\alpha \left [ D_{\alpha \beta \mu \nu}(T^{\alpha \beta}F^{n-1}) - \frac{1}{2}D_{\mu \nu} \left ( TF^{n-1} \right ) \right ] + \Omega ^{\alpha \beta}\frac{\delta T_{\alpha \beta}}{\delta g^{\mu \nu}} - g_{\mu \nu} \frac{V(\phi)}{2} \, ,
\label{gmunu}
\end{eqnarray}
where we have defined the following differential operators:
\begin{eqnarray}
D_{\alpha \beta \mu \nu} & \equiv & \frac{1}{4} \left [ \left ( g_{\mu \alpha}g_{\nu \beta} + g_{\nu \alpha}g_{\mu \beta} \right ) \Box + g_{\mu \nu} \left ( \nabla_{\alpha}\nabla_{\beta}+\nabla_{\beta}\nabla_{\alpha} \right ) - \left ( g_{\mu \alpha}\nabla_{\beta}\nabla_{\nu} + g_{\nu \alpha}\nabla_{\beta}\nabla_{\mu} + g_{\mu \beta} \nabla_{\alpha}\nabla_{\nu} + g_{\nu \beta}\nabla_{\alpha}\nabla_{\mu} \right ) \right ] \, , \nonumber \\
D_{\mu \nu} & \equiv & g_{\mu \nu}\Box - \frac{1}{2} \left ( \nabla_{\mu}\nabla_{\nu}+\nabla_{\nu}\nabla_{\mu} \right ) \, .
\end{eqnarray}
Here, $\Box\equiv \nabla^i\nabla_i$ is the d'Alambertian operator.  In Eq.(\ref{gmunu}), $\Omega_{\mu \nu}$ is a tensor that will not play any role if $T_{\mu \nu}$ does not have any metric dependence, as in the case we are considering: for the purpose of the ensuing discussions, we will omit it. The trace of Eq.(\ref{gmunu}) reads:
\begin{eqnarray}
- R + 4\Lambda - \frac{\lambda}{2}T = 2\alpha F ^n(n-1) + \frac{n\alpha}{2} \left ( g _{\mu \nu}\Box + \nabla _{\mu}\nabla _{\nu} + \nabla _{\nu}\nabla _{\mu} \right ) \left ( T ^{\mu \nu}F ^{n-1} \right ) - \frac{3n\alpha}{2}\Box \left ( TF ^{n-1} \right ) - 2V(\phi) \, .
 \label{trace}
\end{eqnarray}
As we will demonstrate further on, the case where $n=1$ is particularly interesting given that, as anticipated, it is equivalent to a specific case of a mimetic Horndeski model.

\section{Cosmological solutions}
\label{solutions}

In what follows, we shall only consider the flat Friedmann-Lema\^{i}tre-Robertson-Walker (FLRW) metric, whose line element is given by:
\begin{eqnarray}
ds ^2 = -dt ^2 + a ^2(t) \delta_{ij}dx ^idx ^j \, ,
\label{flrw}
\end{eqnarray}
where $a\equiv a(t)$ is the scale factor, and $i,j=1,2,3$. Taking the hypersurfaces of constant time to be equal to those of constant $\phi$, and making use of the constraint on the gradient of the scalar field given by Eq.(\ref{multiplierequation}), we see that the field can be identified (up to an integration constant) with time:
\begin{eqnarray}
\phi = t \, ,
\end{eqnarray}
and thus $\nabla _{i}\phi = (\dot{\phi},0,0,0)$. In this case, the only non-vanishing component of the tensor $T_{\mu \nu}$ is the $(0,0)$ one, i.e.:
\begin{eqnarray}
T _{00} = \dot{\phi} ^2 = 1 \, , \quad T_{0i} = T_{i0} = T_{ij}=0\, , \quad i,j=1,2,3\,.
\end{eqnarray}
We can now choose two independent equations to study the system, and our choice will fall upon Eqs.(\ref{equationscalar},\ref{trace}). We evaluate the trace equation [Eq.(\ref{trace})], which considerably simplifies given that all quantities now depend exclusively on time. The calculation is laborious but relatively straightforward and will be sketched in Appendix A. The final equation we obtain is: 
\begin{eqnarray}
\frac{\lambda}{2} = 6\dot{H} + 12H ^2 - 4\Lambda + \alpha (5n-2)(3H ^2) ^n + 3 ^nn\alpha (2n-1)H ^{2n-2}\dot{H} - 2V(\phi)\,.
\label{tracef}
\end{eqnarray}
From Eq.(\ref{equationscalar}) instead we obtain:
\begin{eqnarray}
\frac{1}{a ^3}\partial _0 \left[ a ^3 \left ( 2n\alpha (3H ^2) ^n - \lambda  \right ) \right] = V'(\phi) \,.
\label{consn}
\end{eqnarray}
The two above equations are of course equivalent to the two equations of motion derived from (\ref{gmunu}) on an FLRW metric, namely:
\begin{eqnarray}
0 & = & \Lambda - 3H^2 + \frac{\alpha}{2}(1-4n)(3 H^2)^n + \frac{\lambda}{2} + \frac{V(\phi)}{2} \, , \\
0 & = & \Lambda - 3H^2 - 2\dot H + \frac{\alpha}{2}(1-2n)(3 H^2)^n + 3^{n-1}\alpha n(1-2n)\dot H H^{2n-2} + \frac{V(\phi)}{2} \, .
\label{m}
\end{eqnarray}
A final remark is in order. For $V(\phi)=0$, from Eq.(\ref{consn}) one obtains:
\begin{eqnarray}
\left [ 2n\alpha(3H^2)^n - \lambda \right ] = \frac{C_0}{a^3} \, ,
\label{friedmann}
\end{eqnarray}
where $C_0$ is an integration constant. One can interpret Eq.(\ref{friedmann}) as a generalized Friedmann equation, with $C_0$ setting the amount of mimetic dark matter in the Universe (in fact, the corresponding energy density scales as $a ^{-3}$, as is expected for cold dark matter). It is also clear that for $\alpha=0$ one recovers the result obtained for mimetic gravity in \cite{mukhanov1,mukhanov2}. \\

The appearance of an effective collisionless cold dark matter component as an integration constant of the equations of motion should not entirely come as a surprise. Recall our model is related to the projectable version of Ho\v{r}ava-Lifshitz gravity, where the lapse is a function of time only, and not spatial coordinates. This implies that the Hamiltonian constraint, which is derived by varying the action with respect to the lapse, is not a local equation to be satisfied at each space-time point, but rather, an equation which is integrated over a constant time hypersurface. Because an integrated condition is less restrictive than an analogous local one, the class of solutions which satisfy the Hamiltonian constraint is richer. It was shown in \cite{mukohyama1} that the relevant set of constraints (Hamiltonian and momentum) lead to the appearance of an integration constant which can be viewed as a pressureless contribution to the energy-momentum tensor. Just as in mimetic gravity, this pressureless component can be identified with collisionless cold dark matter, which thus emerges once more as a purely geometrical effect. Notice that the appearance of an effective pressureless dark matter fluid in the projectable version of Ho\v{r}ava-Lifshitz gravity has been identified in \cite{kocharyan,blas,kobakhidze} as well. \\

\subsection{The $n=1$ case}

Let us consider a more specific setting of the mimetic CRG-like model [Eq.(\ref{cognolaa})] where we set $n=1$ (and thus $\alpha$ is dimensionless). We take $\Lambda=0$, given that the cosmological constant can be incorporated in the potential. In this case Eqs.(\ref{tracef},\ref{consn}) lead to:
\begin{eqnarray}
\lambda & = & 6(2+\alpha)\dot{H} + 6 ( 4+3\alpha )H^2-4V(\phi) \, , \\
\label{t1}
V'(\phi) & = &\frac{1}{a ^3}\partial _0 \left[ a ^3 \left (6\alpha H ^2 - \lambda   \right ) \right] \, .
\label{l}
\end{eqnarray}
Further, note that the derivative with respect to $\phi$ is equal to the time derivative (provided the hypersurfaces of $\phi$ have been appropriately chosen), therefore Eq.(\ref{l}) becomes:
\begin{eqnarray}
\frac{1}{a ^3}\partial _t \left [ a ^3 \left ( 6\alpha H ^2 - \lambda \right ) \right ] = \frac{\partial V}{\partial t} \, .
\label{integrate}
\end{eqnarray}
Note also that, for $n=1$ and  $\Lambda=0$, one obtains from Eq.(\ref{m}):
\begin{eqnarray}
 2\dot{H}+3 H^2=\frac{V}{(2+\alpha)} \, .
\label{z}
\end{eqnarray}
This once more demonstrates that the model we are considering on the FLRW metric is essentially equivalent to the model proposed in \cite{mukhanov2} (see also \cite{rabochaya}) in the limit where $\alpha \rightarrow 0$. \\

Furthermore, for consistency, it it easy to show that Eq.(\ref{l}) is a consequence of Eqs.(\ref{t1},\ref{z}). Thus, one may choose to deal only with (\ref{z}). This equation is a non-linear Riccati type equation. It is a well known fact that it may be transformed in a linear second order differential equation by means of the Sturm-Liouville canonical substitution:
\begin{eqnarray}
H = \frac{2}{3}\frac{\dot{u}}{u} \, .
\end{eqnarray}
After performing this substitution, we are left with:
\begin{eqnarray}
a(t) = u^{2/3} \, ,
\end{eqnarray}
from which we easily derive:
\begin{eqnarray}
\ddot{u} - \frac{3}{4(2+\alpha)}V u=0 \, .
\label{z1}
\end{eqnarray}
We can use Eq.(\ref{z1}) as a reconstruction equation and as starting point for discussing a number of examples. We begin by noting that, if $V$ is a constant, one recovers the de Sitter solution with $u\sim \exp[H_0 t]$, $H_0$ being a constant Hubble parameter.  \\

Another quite natural choice for the potential is a quadratic one, i.e.:
\begin{eqnarray}
V(\phi)=3(2+\alpha)\left [ H_0^2+\beta ^2 \left ( 2\phi-\phi_0\right)\left(-H_0+\frac{\beta^2}{4}(2\phi-\phi_0) \right ) - \frac{2}{3}\beta ^2  \right] \, ,
\end{eqnarray}
where $H_0$ is a fixed Hubble parameter and $\beta \, ,\phi_0$ are dimensional constants, with dimensions  $[\beta]=[\phi_0^{-1}]=[H]$. Given that $\phi=t$, we find that the exact solution for $u$ reads:
\begin{eqnarray}
u(t)=u_0 e^{\frac{3}{2}H_0 t-\frac{3\beta^2}{4} t(t-2 t_0)}\,,\quad t_0\equiv\frac{\phi_0}{2} \, ,
\end{eqnarray}
where $u_0$ is a constant and $t_0$ is a fixed time. This solution can be interpreted as describing a Starobinsky-like inflationary epoch \cite{starobinsky} in the Jordan frame. The Hubble parameter can be derived as:
\begin{eqnarray}
H \equiv \frac{2}{3}\frac{\dot{u}}{u} = H_0 - \beta ^2 \left ( t-t_0\right ) \, .
\label{starobinsky}
\end{eqnarray}
From Eq.(\ref{starobinsky}) one sees that, for $t$ close to $t_0$, one has a quasi-de Sitter expansion. On the other hand, for large $t_0 \ll t$, $H$ approaches zero. \\

We provide another example starting from the following choice:
\begin{eqnarray}
V(\phi) = \frac{4 A^2(2+\alpha)}{3}\frac{\cosh A \phi}{1+\cosh A \phi} \, ,\, ,
\end{eqnarray}
where $A$ is a constant with mass-dimension 1. The exact solution is readily found, and is given by:
\begin{eqnarray}
u(t) = 1+\cosh A t
\end{eqnarray}
In terms of $a(t)$, this solution represents a cosmological bounce (see e.g. \cite{battlefeld} for a recent review on bounce cosmologies) with Hubble parameter 
\begin{equation}
H=\frac{2A}{3}\frac{\sinh At}{(1+\cosh A t )}\,,\quad -\infty<t<+\infty\,.\label{bouncesol}
\end{equation}
 This example shows that the mimetic fluid may act as a phantom fluid.   

\subsection{Perturbations around the FLRW metric}

In order to investigate and ascertain the cosmological viability of the $n=1$ CRG-like model, one has to consider perturbations around the flat FLRW metric, Eq.(\ref{flrw}). We consider only scalar perturbations, given that vector perturbations are not produced in the most common models of inflation and quickly decay with the expansion of the Universe, and we are not interested in primordial tensor modes. Thus, it is simplest for us to work in conformal Newtonian gauge, where the line element reads:
\begin{eqnarray}
ds^2 = -( 1+2\Phi(t, {\bf x}) )dt^2 + a^2(t)( 1-2\Psi(t, {\bf x}) )\delta_{ij}dx^i dx^j \, , \quad i,j = 1,2,3 \, ,
\label{NG}
\end{eqnarray}
where $\Phi(t, {\bf x})$ and $\Psi(t, {\bf x})$ are functions of the space-time coordinates and $|\Phi(t,x)\,,\Psi(t,x)|\ll 1$. Thus, to lowest order, $g^{00}(t,x)\simeq-1+2\Phi(t,x)$ and $g^{11}(t,x)\simeq a(t)^{-2}(1+2\Psi(t,x))$. We perturb the mimetic field as:
\begin{eqnarray}
\phi = t+ \delta\phi (t,x) \, ,
\end{eqnarray}
$\phi(t,x)$ being a function of the space-time coordinates. The mimetic constraint, Eq.(\ref{multiplierequation}), implies the following relation:
\begin{eqnarray}
\delta \dot{\phi}(t,x) = \Phi(t,x) \, .
\label{phiPhi}
\end{eqnarray}
The following relations hold true as well:
\begin{eqnarray}
T_{00} = 1 + 2\delta \dot{\phi} \, , \quad T_{0i} = \partial_ i \delta\phi \, , \quad T = -1 + {\cal O}(\Phi(t,x)^2) \, .
\end{eqnarray}
From the $(i,j)$ components of (\ref{gmunu}), when $i,j=1,2,3 \, , i\neq j$, we obtain to first order in $\delta\phi(t,x)$ [we can use the (1-2)-component]:
\begin{eqnarray}
G_{12} \left ( 1 - \frac{\alpha}{2} \right ) = \alpha D_{\alpha \beta 12}T^{\alpha \beta} \, ,
\end{eqnarray}
where:
\begin{eqnarray}
G_{12} = -\partial _x\partial _y ( \Phi - \Psi ) \, , \quad D_{\alpha \beta 12}T^{\alpha \beta} = H\partial _x\partial _y \delta \phi + \partial _x\partial _y \delta \dot{\phi} \, .
\end{eqnarray}
It then follows that:
\begin{eqnarray}
\Psi=\Phi+\left(\frac{2\alpha}{2-\alpha}\right)(H\delta\phi+\delta\dot\phi)\,.
\end{eqnarray}
The (0,1)-component of Eq.(\ref{actionpotential}) reads:
\begin{eqnarray}
G_{01} \left ( 1 + \frac{\alpha}{2} \right ) = \alpha \dot{H} \partial _x\delta\phi + \frac{\lambda}{2} \partial _x\delta\phi + \alpha D_{\alpha \beta 01}T^{\alpha \beta} \, ,
\end{eqnarray}
where:
\begin{eqnarray}
G_{01}=2\partial _x \left ( \dot\Psi + H\Phi \right ) \, , \quad D_{\alpha \beta 01}T^{\alpha \beta} = -(H^2 + \dot{H}) \partial _x\delta\phi \, , \quad \lambda = 6\alpha H^2 - 4\dot{H} - 2\alpha\dot{H} \, .
\end{eqnarray}
Thus, one arrives at:
\begin{eqnarray}
\delta \ddot{\phi} + H \delta \dot{\phi} + \dot{H} \delta\phi = 0 \, .
\end{eqnarray}
Some remarks are in order at this point. First, the perturbation equation is equal to the one obtained in \cite{mukhanov2}. Furthermore, as noticed there, the sound speed is vanishing. As a consequence, it is not possible to define the quantum fluctuations of the mimetic field as in the standard inflation models. In order to overcome this drawback, one has to modify the model. In our case, one could investigate the $n=2$ case, which will be done elsewhere. Alternatively, one can try to modify the model along the lines of \cite{mukhanov2,capela}.

\section{Modified higher order mimetic model}
\label{modified} 

In order to modify our original $n=1$ CRG-like mimetic model, we first recall the identity \cite{kobayashi}:
\begin{eqnarray}
-\frac{1}{2}g^{\mu \nu}\nabla_{\mu}\phi\nabla_{\nu}\phi R + (\Box \phi)^2 - (\nabla_{\mu} \nabla_{\nu}\phi)^2 = G^{\mu \nu}\nabla_{\mu}\phi\nabla_{\nu}\phi + \mbox{total derivative} \, .
\end{eqnarray}
and rewrite the  $n=1$ action [Eq.(\ref{actionpotential})] in the form:
\begin{eqnarray}
S = \frac{1}{2}\int d ^4x\sqrt{-g} \left [ R(1 +\frac{\alpha}{2}g^{\mu \nu}\nabla_{\mu}\phi\nabla_{\nu}\phi) - \alpha(\Box \phi)^2 + \alpha (\nabla_{\mu} \nabla_{\nu}\phi)^2 - \frac{\lambda}{2} \left ( g ^{\mu \nu}\nabla _{\mu}\phi\nabla _{\nu}\phi + 1 \right ) - V(\phi) \right ] \,.
\label{actionpotential1}
\end{eqnarray}
In this form, the above action is still of the mimetic Horndeski form (see e.g. \cite{arroja,rabochaya,myrzakulov10}). We now modify the model as follows:
\begin{eqnarray}
S = \frac{1}{2}\int d ^4x\sqrt{-g} \left [ R(1 + a g^{\mu \nu}\nabla_{\mu}\phi\nabla_{\nu}\phi) - \frac{c}{2}(\Box \phi)^2 + \frac{b}{2} (\nabla_{\mu}\nabla_{\nu}\phi)^2 - \frac{\lambda}{2} \left ( g ^{\mu \nu}\nabla _{\mu}\phi\nabla _{\nu}\phi + 1 \right )  - V (\phi) \right ] \,.
\label{actionpotential2}
\end{eqnarray}
In this form, and without the mimetic constraint, the action describes a higher order derivative model in the scalar sector, namely with equations of motion which are of fourth order. The presence of the mimetic constraint renders the potential instability problem milder. Our original $n=1$ model is recovered when $a=\frac{\alpha}{2}$ and $b=c=2 \alpha$. Thus we can interpret $(b-c)$ as a ``Horndeski breaking parameter". \\

A comment is in order here. By reading off the extrinsic curvature on spatial slices from Eq.(\ref{actionpotential1}), one immediately notices that the theory can be related to the $\lambda = 1$ version of projectable Ho\v{r}ava-Lifshitz gravity. Recall that in projectable Ho\v{r}ava-Lifshitz gravity an extra scalar mode emerges, because of differences in the diffeomorphism structure with respect to General Relativity. The existence of this mode has been discussed at length in the literature on HLG, see e.g. \cite{horava,horava1,cai,wang,tang}. In particular, \cite{cai} showed how, in the IR and for $\lambda = 1$, the would-be dynamical scalar mode is actually non-propagating. This of course explains why the original CRG-like model is free from extra propagating scalar degrees of freedom: the Horndeski condition $b = c = 2\alpha$ corresponds precisely to the condition that $\lambda = 1$ in projectable Ho\v{r}ava-Lifshitz gravity. Moreover, this should not come as a surprise given that recent work has shown how the sound speed of scalar perturbations in mimetic Horndeski model, within which our CRG-like model falls, is identically zero. For the same reason, we expect that the higher order modification given by Eq.(\ref{actionpotential2}) violates the $\lambda = 1$ condition, thus giving rise to a non-trivial dispersion relation for the scalar mode which is now dynamical. In fact, one could work backwards and realize that violating the $\lambda = 1$ condition or equivalently $b = c = 2\alpha$ is necessary in order to have a non-vanishing sound speed. We leave further exploration of these very interesting connections to future work. \\

Let us study the equations of motion of the modified higher order mimetic CRG-like model. The equations of motion for the gravitational field, implemented by the mimetic constraint, read:
\begin{eqnarray}
(1-a)G_{\mu \nu} & = & \frac{1}{2}g_{\mu \nu} \left [ \frac{b}{2}\phi^{\alpha \beta}\phi_{\alpha \beta} - \frac{c}{2}(\Box\phi)^2 -V(\phi) \right ] + \lambda \nabla_{\mu}\phi\nabla_{\nu}\phi \nonumber \\
& - & b \phi_{\mu \rho}\phi^{\rho}_{\nu} + \frac{b}{2}g^{\alpha \beta} \left [ \nabla_{\alpha}(\phi_{\mu \nu} \nabla_{\beta}\phi) - \nabla_{\alpha}(\phi_{\mu \beta}\nabla_{\nu}\phi) - \nabla_{\alpha}(\phi_{\nu \beta}\nabla_{\mu}\phi) \right ] \nonumber \\
& + & c \left [ \phi_{\mu \nu} + g_{\mu \nu}g^{\alpha \beta}\nabla_{\alpha}(\Box\phi\nabla_{\beta}\phi) - \nabla_{\mu} \Box\phi \nabla_{\nu}\phi - \nabla_{\nu}\Box\phi \nabla_{\mu}\phi \right ] \, ,
\end{eqnarray}
where $\phi_{\alpha \beta} \equiv \nabla_{\alpha}\nabla_{\beta}\phi$. On a flat FLRW spacetime, the relevant bulk equation reads:
\begin{eqnarray}
2\dot{H} + 3H^2 = c_v V(\phi) \, ,
\label{bulk}
\end{eqnarray}
where we have defined $c_v$ as follows:
\begin{eqnarray}
c_v \equiv \frac{2}{4 - 4a - b + 3c} \, . 
\end{eqnarray}
Again, as a consistency check, we verify that by setting $b = c = 2 \alpha$ and $a = \frac{\alpha}{2}$ such that $c_v=1/(2+\alpha)$, we recover the previous results. Concerning the bulk equation [Eq.(\ref{bulk})], we once again introduce the auxiliary function $u$ as:
\begin{eqnarray}
H = \frac{2}{3}\frac{\dot{u}}{u} \, ,
\end{eqnarray}
from which one has:
\begin{eqnarray}
a(t) = u^{\frac{2}{3}} \, ,
\end{eqnarray}
and:
\begin{eqnarray}
\ddot{u} - \frac{3 c_v V(\phi)}{4} u=0\,.
\label{z2}
\end{eqnarray}
One can use Eq.(\ref{z2}) to obtain exact solutions as in Section \ref{solutions}. For instance, making the following choice:
\begin{eqnarray}
V(\phi) = \frac{4 A^2}{3c_v}\frac{\cosh A \phi}{1+\cosh A \phi} \, ,
\end{eqnarray}
one recovers the bounce solution
\begin{eqnarray}
u(t) = 1+\cosh A t \, .
\end{eqnarray}
Alternatively, along the lines of \cite{rabochaya}, one may introduce the the e-fold time $N \equiv -\ln a$. As a result, the equation of motion for $H$ becomes:
\begin{eqnarray}
-\frac{dH^2}{dN} + 3 H^2 = c_vV(N) \, .
\label{r}
\end{eqnarray}
The general solution reads:
\begin{eqnarray}
H^2(N) = \text{e}^{3N}\left ( C - \int dN \text{e}^{-3N} c_v V(N) \right) \, .
\label{r1}
\end{eqnarray}
The term depending on the constant $C$ plays the role of cosmological dark matter. Other exact solutions can be obtained with a suitable choice of the potential. For example, by making the choice:
\begin{eqnarray}
V(N)=V_0 N\,,
\end{eqnarray}
$V_0$ being a positive constant, one finds:
\begin{eqnarray}
H^2(N) = \text{e}^{3N}C + \frac{V_0 c_v}{3} \left ( N+\frac{1}{3} \right ) \, .
\end{eqnarray}
Thus, if the condition $c_v>0$ is met, one can describe inflation when $1\ll N$ (see e.g. \cite{myrzakulov10}).

\subsection{Perturbations of the modified higher order mimetic model}

Let us now discuss perturbations in the modified higher order mimetic CRG-like model. We have seen that in the FLRW space-time bulk, the exact cosmological solutions for the modified higher order mimetic model are similar to the ones of the Horndeski mimetic model we are interested in. We now turn to the study of cosmological perturbations. We work once more in the comoving Newtonian gauge, with line element defined by Eq.(\ref{NG}). Thus, as before $\delta \dot{\phi}=\Phi$. The spatial components of the perturbed equations give:
\begin{eqnarray}
\Psi=\Phi+\frac{b}{2-2\alpha}\left(\delta \dot{\phi}+H\delta\phi \right)\,.
\label{r2}
\end{eqnarray}
Finally, the $_{0i}$ and $_{i0}$ components of the perturbed equations give:
\begin{eqnarray}
\delta \ddot{\phi} + H \delta \dot{\phi}-\frac{c_s^2}{a^2}\nabla^2 \delta \phi + \dot{H}+ \delta\phi=0\,,
\end{eqnarray}
where $c_2$ is defined as:
\begin{eqnarray}
c_2 \equiv \frac{(2a-2-b)(4+3c-4a-b)}{4(a-1)} \, ,
\end{eqnarray}
and the non-vanishing squared sound speed, $c^2_s$, reads:
\begin{eqnarray}
c^2_s \equiv \frac{b-c}{2 c_2} \, .
\end{eqnarray}
The result is completely analogous to that obtained in \cite{mukhanov2,capela}, albeit with a modified sound speed. As expected, we recover the result presented in \cite{mukhanov2} when we set $a=0$, $b=0$. Moreover, as expected, the sound speed vanishes in the Horndeski case, when $b = c$, and is proportional to the Horndeski breaking parameter $b-c$.

\section{Conclusions}
\label{conclusion}

In this paper, we have explored connections between two modified gravity frameworks. The first is a variant of the Nojiri-Odintsov covariant renormalizable gravity, a fully diffeomorphism invariant Ho\v{r}ava-like theory of gravity which breaks diffeomorphism invariance dynamically, by means of a non-standard coupling to a perfect fluid. The second is mimetic gravity, which by means of a scalar field (the mimetic field) constrained by a Lagrange multiplier, contains a similar fluid. We have extended the Ho\v{r}ava-like theory to include the mimetic field, and have provided a first inspection of the consequences of this extension. \\

We have thus considered a generalization of the mimetic covariant Ho\v{r}ava-like model. For the case $n=1$, we have shown that there subsists an equivalence with a particularly simple class of Horndeski models. As in the case of the original mimetic gravity proposal of \cite{mukhanov1,mukhanov2}, we have shown that with the addition of a suitable potential it is possible to reconstruct several viable cosmological scenarios, by using Eq.(\ref{z1}). As an example, we have shown how to realize a Starobinsky-like inflationary epoch and a bounce solution, starting from well-motivated potentials. \\

By studying perturbations of the $n=1$ model around a flat FLRW background, we have shown that the fluid in question behaves as an irrotational fluid with vanishing sound speed, which renders the usual definition of quantum perturbations problematic. Parallel work has shown how this is the case in the most general mimetic Horndeski model \cite{bartolo}. Therefore we address this problem by going beyond the Horndeski form, modifying the action as in Eq.(\ref{actionpotential2}). When we consider perturbations in this model around a flat FLRW background, the resulting sound speed is no longer zero, but proportional to the Horndeski breaking parameter. Thus perturbations which will then grow under gravitational instability to seed the large-scale structure in the Universe can be defined sensibly in this model. \\

In conclusion, we have modified a variant of the Nojiri-Odintsov covariant Ho\v{r}ava-like model. The resulting model is thus a theory of gravity which preserves diffeomorphism invariance at the level of the action, reduces to General Relativity in the infrared, allows the realization of a number of interesting cosmological scenarios, is well defined when considering perturbations around a flat FLRW background, and features cosmological dark matter emerging as an integration constant. We have also commented on relations between this theory and projectable Ho\v{r}ava-Lifshitz gravity. \\

There are several avenues for further work on this topic. Firstly, it is necessary to study the full implications of the likely loss of renormalizability when a non-constant potential for the mimetic field is included, as well as solutions to this important problem. Another interesting question concerns the size of the Horndeski breaking parameter, and hence of the sound speed. Depending on the magnitude of this quantity, one could in principle use our model to address some of the outstanding problems of collisionless cold dark matter on small scales, by appropriately suppressing small-scale power in the matter power spectrum. In the future, it would be interesting to consider the $n \geq 2$ models, together with further interesting cosmological solutions. In particular it would be certainly important to study the evolution of the gravitational potential and the form of the resulting late-time matter power spectrum. We reserve the study of these issues for future work.

\section*{Acknowledgements}

SV is supported by the Swedish Research Council (VR) through the Oskar Klein Centre. We are indebted to the referees for valuable and detailed suggestions which greatly improved the quality of our work. We are very grateful to Sergei Odintsov for useful discussions and comments on a draft of the manuscript. We also thank Fred Arroja, Nicola Bartolo, Purnendu Karmakar and Sabino Matarrese for useful discussions and for sharing ideas of their work \cite{bartolo} prior to publication.

\section*{Appendix A}

Here we sketch the derivation of Eq.(\ref{tracef}). To begin we evaluate $F$:
\begin{eqnarray}
F = T _{\mu \nu}R ^{\mu \nu} - \frac{RT}{2} = T ^{00}R _{00} - \frac{RT}{2} = 3H ^2 \ ,
\end{eqnarray}
where we have used the fact that the only non-zero component of $T _{ij}$ is $T _{00}$, and in FRW $R _{00} = -3\ddot{a}/a$, $R = 6[\ddot{a}/a + (\dot{a}/a) ^2]$. Next, we evaluate $\Box (TF ^{n-1})$, where $T = g ^{\mu \nu}T _{\mu \nu} = -1$:
\begin{eqnarray}
\Box (TF ^{n-1}) & = & 3 ^{n-1} \left ( \frac{\partial ^2}{\partial t ^2} + 3H\frac{\partial}{\partial t} \right )H ^{2n-2} \\ \nonumber
& = & 3 ^{n-1}(2n-2)(2n-3)\dot{H} ^2H ^{2n-4} + 3 ^{n-1}(2n-2)\ddot{H}H ^{2n-3} + 3 ^n(2n-2)\dot{H}H ^{2n-2}
\end{eqnarray}
Next, we evaluate $g _{\mu \nu}\Box (T ^{\mu \nu}F ^{n-1})$. Metric compatibility can be used to argue that this term is actually equal to the one we just calculated:
\begin{eqnarray}
g _{\mu \nu}\Box (T ^{\mu \nu}F ^{n-1}) & = & \Box (TF ^{n-1}) \\ \nonumber
& = & 3 ^{n-1}(2n-2)(2n-3)\dot{H} ^2H ^{2n-4} + 3 ^{n-1}(2n-2)\ddot{H}H ^{2n-3} + 3 ^n(2n-2)\dot{H}H ^{2n-2}
\end{eqnarray}
Obviously, $(\nabla _{\mu}\nabla _{\nu} + \nabla _{\nu}\nabla _{\mu})(T ^{\mu \nu}F ^{n-1}) = 2\nabla _{\mu}\nabla _{\nu}(T ^{\mu \nu}F ^{n-1})$, so we only need to evaluate $\nabla _{\mu}\nabla _{\nu}(T ^{\mu \nu}F ^{n-1})$. We first evaluate $\nabla _{\nu}(T ^{\mu \nu}F ^{n-1})$:
\begin{eqnarray}
\nabla _{\nu}(T ^{\mu \nu}F ^{n-1}) = F ^{n-1}(\partial _{\nu}T ^{\mu \nu} + \Gamma ^{\mu} _{\beta \sigma}T ^{\sigma \beta} + \Gamma ^{\nu} _{\nu \sigma} T ^{\mu \sigma}) + T ^{\mu \nu}\partial _{\nu}(F ^{n-1}) \equiv T ^{\mu}
\label{ta}
\end{eqnarray}
Therefore:
\begin{eqnarray}
\nabla _{\mu}\nabla _{\nu}(T ^{\mu \nu}F ^{n-1}) = \nabla _{\mu}T ^{\mu} = \frac{1}{\sqrt{-g}}\partial _{\mu}(\sqrt{-g}T ^{\mu}) = \partial _{\mu}T ^{\mu} + \frac{3}{a}T ^{\mu}\partial _{\mu}a
\end{eqnarray}
By making use of Eq.(\ref{ta}) and expanding, we get:
\begin{eqnarray}
& & \nabla _{\mu}\nabla _{\nu}(T ^{\mu \nu}F ^{n-1}) = \partial _{\mu}\left [ \nabla _{\nu}(T ^{\mu \nu}F ^{n-1}) \right ] + \frac{3}{a}\left [ \nabla _{\nu}(T ^{\mu \nu}F ^{n-1}) \right ]\partial _{\mu}a = \\ \nonumber & & (\partial _{\nu}T ^{\mu \nu} + \Gamma ^{\mu} _{\beta \sigma}T ^{\sigma \beta} + \Gamma ^{\nu} _{\nu \sigma}T ^{\mu \sigma})\partial _{\mu}(F ^{n-1}) + F ^{n-1}(\partial _{\mu}\partial _{\nu}T ^{\mu \nu} + T ^{\sigma \beta}\partial _{\mu}\Gamma ^{\mu} _{\beta \sigma} + \Gamma ^{\mu} _{\beta \sigma}\partial _{\mu}T ^{\sigma \beta} + T ^{\mu \sigma}\partial _{\mu}\Gamma ^{\nu} _{\nu \sigma} + \Gamma ^{\nu} _{\nu \sigma}\partial _{\mu}T ^{\mu \sigma}) \\ \nonumber
& & + \partial _{\mu}\partial _{\nu}(F ^{n-1})T ^{\mu \nu} + \partial _{\mu}(T ^{\mu \nu})\partial _{\nu}(F ^{n-1}) + \frac{3}{a}F ^{n-1}(\partial _{\nu}T ^{\mu \nu} + \Gamma ^{\mu} _{\beta \sigma}T ^{\sigma \beta} + \Gamma ^{\nu} _{\nu \sigma}T ^{\mu \sigma})\partial _{\mu}a + \frac{3}{a}T ^{\mu \nu}\partial _{\nu}(F ^{n-1})\partial _{\mu}a
\label{nablaalpha}
\end{eqnarray}
Of the fourteen terms in Eq.(\ref{nablaalpha}), only five are nonzero. This can easily be shown by repeatedly using the fact that $T ^{\mu \nu} = 0$ if $\mu , \nu \neq 0$, $\partial _{\mu}T ^{\alpha \beta} = 0$ and $\Gamma ^{0} _{00} = \Gamma ^{i} _{00} = 0$. The only five nonzero terms are:
\begin{eqnarray}
\Gamma ^{\nu} _{\nu \sigma}T ^{\mu \sigma}\partial _{\mu}F^{n-1} & = & 3 ^n(2n-2)\dot{H}H ^{2n-2} \\
\partial _{\mu}\partial _{\nu}(F ^{n-1})T ^{\mu \nu} & = & 3 ^{n-1}\left [ (2n-2)(2n-3)\dot{H} ^2H ^{2n-4} + (2n-2)\ddot{H}H ^{2n-3} \right ] \\
F ^{n-1}\partial _{\mu}(\Gamma ^{\nu} _{\nu \sigma})T ^{\mu \sigma} & = & 3 ^{n+1}H ^{2n} + 3 ^n(4n-3)\dot{H}H ^{2n-2} + 3 ^{n-1} \left [ (2n-2)(2n-3)\dot{H} ^2H ^{2n-4} + (2n-2)\ddot{H}H ^{2n-3} \right ] \\
\frac{3}{a}F ^{n-1}\Gamma ^{\nu} _{\nu \sigma}T ^{\mu \sigma}\partial _{i}a & = & 3(3H ^2) ^n \\
\frac{3}{a}T ^{\mu \nu}\partial _{\nu}(F ^{n-1})\partial _{\mu}a & = & 3 ^n(2n-2)\dot{H}H ^{2n-2}
\end{eqnarray}
It is then easy to see that:
\begin{eqnarray}
(g _{ij}\Box + \nabla _{i}\nabla _{j} + \nabla _{j}\nabla _{i})(T ^{rs}F ^{n-1}) - 3\Box (TF ^{n-1}) = 6(3H ^2) ^n + 3 ^n(4n-2)\dot{H}H ^{2n-2} \ .
\end{eqnarray}
From this one then easily obtains Eq.(\ref{tracef}).

\end{document}